\documentclass[%
 reprint,nofootinbib,
superscriptaddress,
 amsmath,amssymb,
 aps,
floatfix,
]{revtex4-2}

\usepackage{graphicx}
\usepackage{dcolumn}
\usepackage{bm}
\usepackage{float}
\usepackage{booktabs}
\usepackage{setspace}
\usepackage{caption}

\newcommand{\beq}{\begin{equation}}
\newcommand{\eeq}{\end{equation}}
\newcommand{\bea}{\begin{eqnarray}}
\newcommand{\eea}{\end{eqnarray}}
\newcommand{\beqn}{\begin{equation*}}
\newcommand{\eeqn}{\end{equation*}}
\newcommand{\bean}{\begin{eqnarray*}}
\newcommand{\eean}{\end{eqnarray*}}

\begin{document}

\title{A simple supergravity model of inflation constrained with {\it Planck 2018} data}

\author{Gabriel Germ\'an}
\email{e-mail: gabriel@icf.unam.mx}
\affiliation{Instituto de Ciencias F\'{\i}sicas, Universidad Nacional
Aut\'onoma de M\'exico, Cuernavaca, Morelos, 62210, Mexico}
\author{Juan Carlos Hidalgo}
\affiliation{Instituto de Ciencias F\'{\i}sicas, Universidad Nacional Aut\'onoma de M\'exico, Cuernavaca, Morelos, 62210, Mexico}
\author{Francisco X. Linares Cede\~no }
\affiliation{Departamento de F\'isica, DCI,
Campus Le\'on, Universidad de Guanajuato,
37150, Le\'on, Guanajuato, Mexico}
\author{Ariadna Montiel}
\affiliation{Instituto de Ciencias F\'{\i}sicas, Universidad Nacional
Aut\'onoma de M\'exico, Cuernavaca, Morelos, 62210, Mexico}
\author{J. Alberto V\'azquez}
\affiliation{Instituto de Ciencias F\'{\i}sicas, Universidad Nacional
Aut\'onoma de M\'exico, Cuernavaca, Morelos, 62210, Mexico}

\date{\today}

\begin{abstract}
We study a model of inflation based on $\mathcal{N}=1$ supergravity essentially depending 
on one effective parameter. Under a field transformation we show that this model turns out to be equivalent to a previously studied 
supergravity model known to be ruled out with the original choice of the parameter. Such parameter measures the slope of the 
potential at observable scales.  Through a Bayesian parameter estimation, it is shown how this model is compatible with recent 
CMB temperature measurements by {\it Planck 2018} giving rise to a simple, viable, single field model of inflation. 
The tensor to scalar ratio constraint is found 
to be $r_{0.002}<0.065$  with negative running.  We discuss how observables are invariant under 
the field transformation which leaves unaltered the slow-roll parameters. As a consequence the use of one presentation of the 
model or its field-transformed version is purely a matter of convenience.
\end{abstract}

\maketitle
\section{Introduction}

The inflationary paradigm has proved its validity against early universe observables which are detected with ever-higher precision by modern experiments. Most prominently, the latest data release of the Planck Satellite \cite{PLK18} reports, for instance, a spectrum of primordial scalar fluctuations with index $n_s =0.965 \pm 0.004$, perfectly consistent with a standard slow-roll inflation model \cite{Akrami:2018odb}. Several realizations of inflation offer a link between this paradigm and fundamental physics of the early universe and/or a good fit to data, mostly in terms of a single canonical scalar field dubbed the inflaton \cite{Guth:1980zm} (and see \cite{Martin:2013tda} for a plethora of viable models). A common exercise is the search for characteristic signatures to discriminate between distinct models (potentials). The recent constrain to the tensor-to-scalar ratio $r$ \cite{BKP16} rejects several of the models of inflation which feature large values of this parameter (\cite{Akrami:2018odb}, and see \cite{German:2015qjq} for a generic characteristic of inflationary potentials resulting in large $r$). In this context, symmetries of the slow-roll parameters are here employed to group apparently different potentials in a single description and thus reduce the number of models to test by observations. This may provide physical foundation for empirical models. In this paper, we study a model rooted in supergravity which essentially depends on one parameter identified with the slope of the potential during the inflationary era. The value of this parameter is fixed by means of a Bayesian analysis resulting in a viable model. 
We discuss two interpretations of the possible origin of the inflationary epoch suggested by the very location of the potential. 
In the first presentation of the model given by Eq.~(\ref{SuPo1}) and Fig.~\ref{SugraPot2} we naturally interpret inflation 
as a transient phenomenon with the inflaton rolling from high energy and starting inflation somewhere at $\eta=1$ and ending 
when $\eta=-1$ (in this model $\epsilon$ is always less than $|\eta|)$. 
A {\it total} number of $e$-folds can be quantified containing the necessary number of $e$-folds for observable inflation. 
A field transformation can take this potential (without any deformation) 
to the one given by Eq.~(\ref{sugrapot4}) and Fig.~\ref{SugraPot3} making easier to obtain approximated expressions for 
quantities of interest. As discussed long time ago \cite{Ellis:1982dg, *Nanopoulos:1982bv,*Ovrut:1983my,*Holman:1984yj,Ross:1995dq} it is then natural to interpret the second 
presentation of this model as originating from a previous phase transition from a high energy symmetric potential where thermal effects keep the inflaton initially at the origin, afterwards slowly rolling towards its global minimum. 
In both cases the observables are exactly the same because the potential has not been deformed by the field transformation. 
Physically relevant quantities like the spectral index, tensor to scalar ratio and all other observables do not depend on the 
particular value of  $\phi_H$\footnote{ where the subscript $H$ designates values of quantities where perturbations
are produced, some $50$ $-$ $60$ $e$-folds before the end of inflation.} but only when referred to a specific potential. 
The potential an all its even-number of derivatives are unchanged by the operations of translation and reflection but 
odd-number of derivatives of the potential switch sign (because of the $\phi \rightarrow -\phi$ transformation). 
However the slow-roll (SR) parameters remain unchanged since they involve an even number of factors with odd-number of derivatives 
of the potential. As a consequence observables remain the same as shown in the body of the article.
The outline of the paper is as follows: in Section \ref{The model} we provide a brief presentation of the supergravity model by 
writing the K\"ahler potential and the superpotential thus specifying the F-term part of the scalar potential. Section \ref{Slow Roll} 
contains a brief discussion of the slow-roll parameters and their invariance under the field redefinition. In Section \ref{version}  
we choose a convenient version of the model where analytical approximations can be easily done. Section \ref{Planck} contains 
a Bayesian analysis for the estimation of parameter values which best fit the \textit{Planck 2018} data. Finally Section \ref{Conclusions} 
contains a brief discussion of the main results and conclusions.

\section{A supergravity model of inflation} \label{The model} 

\begin{figure}[tb]
\captionsetup{format=plain,justification=centerlast}
\begin{center}
\includegraphics[width=8.5cm]{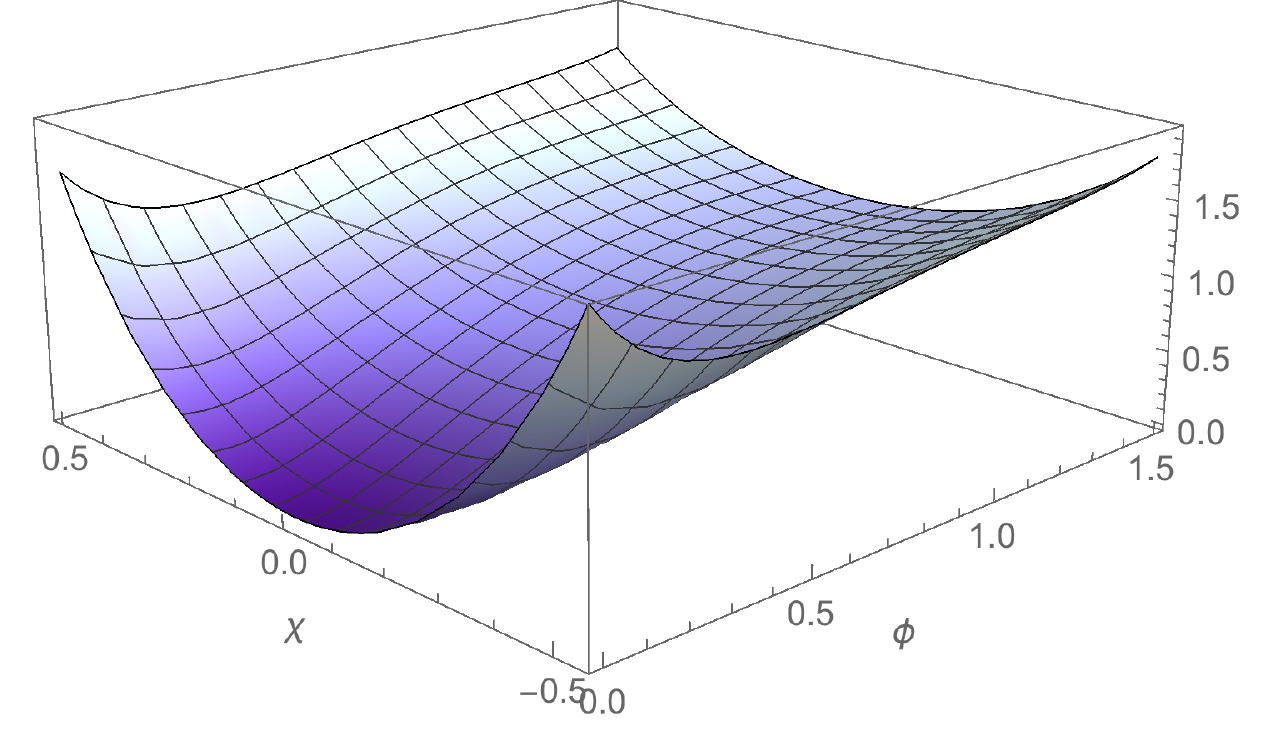}
\caption{\small The full potential given by Eqs.~(\ref{SuPo}) and (\ref{complexfield}) along the $\chi$ and  $\phi$ directions. 
The $\chi$ direction is stable and we can then study the inflationary $\phi$-direction safely on its own. }
\label{SugraPot1}
\end{center}
\end{figure}

The piece of the $\mathcal{N}=1$ supergravity model of interest is given by the action  \cite{Cremmer:1982en}
\beq
\label{sugraction}
I= -\int d^4 x \sqrt {-g}\left(\frac{1}{2}R + G_i^{\,\,j} \partial_{\mu}\Phi^i \partial_{\nu}\Phi^*_jg^{\mu\nu} + V\right),
\eeq
where the K\"ahler metric $G_i^{\,\,j}$ is defined by $G_i^{\,\,j}=\partial^2 G/\left(\partial \Phi^*_j\partial\Phi^i\right)$ and the 
K\"ahler function is $G( \Phi^i, \Phi^*_i)=K( \Phi^i, \Phi^*_i) + \ln |W(\Phi^i)|^2$, $W$ is a holomorphic function of $\Phi^i$ called 
superpotential and $K$ is the K\"ahler potential, a real function depending on the superfields $\Phi^i$ as well as their 
conjugates $ \Phi^*_i$. The $F-term$ part of the scalar potential is given in terms of the real function $G$ as follows
\beq
\label{sugrapot1}
V= e^G\left(G_i(G^{-1})^i_jG^j-3  \right) .
\eeq
In what follows we concentrate in a single chiral superfield $\Phi$ with scalar component $z$.
Thus, the potential is given by
\beq
\label{sugrapot2}
V= e^K\left(F_z (K_{z z^*})^{-1} F^*_{z^*}-3|W|^2  \right),
\eeq
where
\beq
\label{derivative}
F_z \equiv \frac{\partial W}{\partial z}+\frac{\partial K}{\partial z} W, \quad K_{zz^*}\equiv \frac{\partial^2 K}{\partial z\partial z^*} .
\eeq
To first approximation we take the K\"ahler potential to be of the canonical form
\beq
\label{kpot1}
K(z,z^*)= (z-z_0) (z^*-z^*_0)+ \cdot  \cdot  \cdot\, ,
\eeq
with superpotential 
\beq
\label{superpot1}
W(z)=f(z_0) z^2, 
\eeq
where $f(z_0)$ is a constant with dimensions of mass which we simple take as $\Lambda$. The scalar potential becomes
\beq
\label{SuPo}
V=\Lambda^2 e^{|z-z_0|^2} |z |^2 \left( -3 |z |^2+|2+ |z |^2-z_0z^*|^2\right).
\eeq
Writing $z$ in terms of real components
\beq
\label{complexfield}
z=\frac{1}{\sqrt 2}(\phi+i\chi),
\eeq
we find that the $\chi$-direction is a stable direction of the full potential, shown in Fig.~\ref{SugraPot1}.
Thus we set $\chi=\chi_0=0$ and study the potential along the inflationary $\phi$-direction only which is given by
\begin{multline}
\label{SuPo1}
V=\frac{1}{8}\Lambda^2 e^{\frac{1}{2}(\phi-\phi_0)^2}\left( 16 \phi^2 -8\phi_0 \phi^3 + (2+ \phi^2_0)\phi^4  \right. \\
  \left. -2\phi_0 \phi^5 + \phi^6  \right).
\end{multline}

\begin{figure}[tb]
\captionsetup{format=plain,justification=centerlast}
\includegraphics[width=7cm]{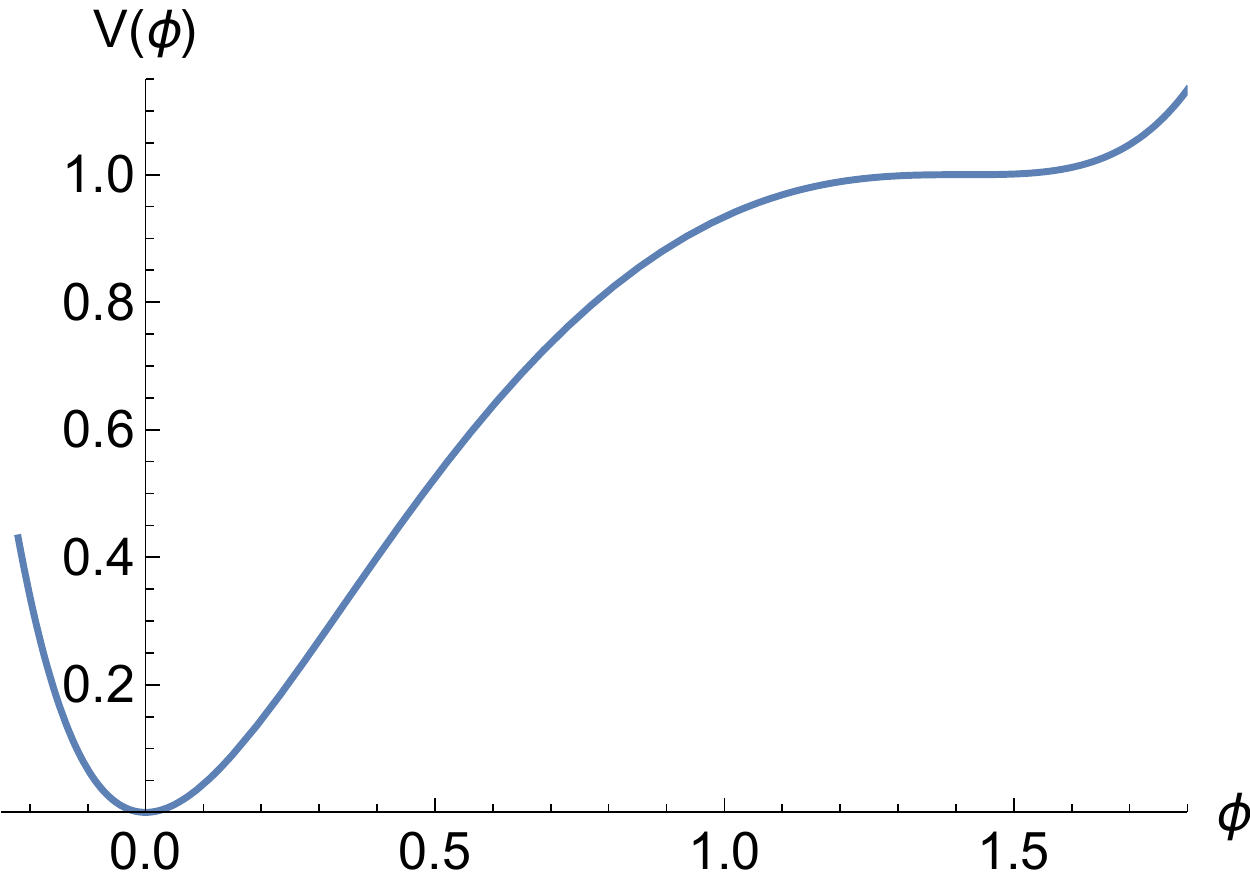}
\caption{\small Schematic plot of the inflationary potential  given by Eq.~(\ref{SuPo1}).   Here, the inflaton rolls towards 
it global minimum (conveniently located at the origin) and inflation is a transient phenomenon with a finite {\it total} number of $e$-folds. 
We can locate the {\it beginning} of inflation at $\eta=1$ and its end when $\eta=-1$ with the parameter $\epsilon$ always 
smaller than $|\eta|.$ A fit to the {\it Planck-2018} data suggest the the total number of $e$-folds is no more than triple the usually 
required  60 $e$-folds of observable inflation. In its form given by Eq.~(\ref{SuPo1}) the resulting potential is somewhat 
reminiscent of Starobinsky's but with a different asymptotic behavior. }
\label{SugraPot2}

\end{figure}

\begin{figure}[tb]
\captionsetup{format=plain,justification=centerlast}
\begin{center}
\includegraphics[width=7cm]{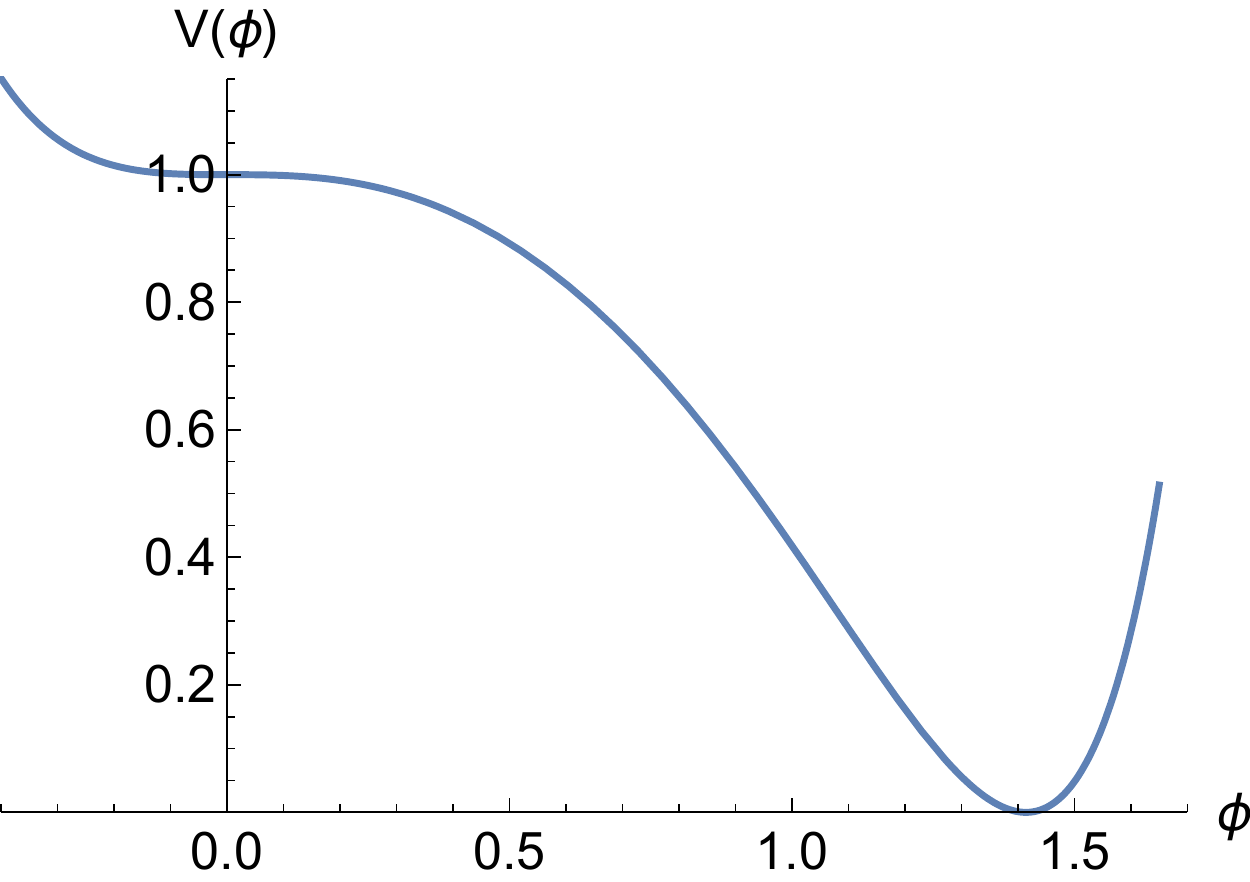}
\caption{\small Schematic plot of the inflationary potential  given by Eq.~(\ref{sugrapot4}).  The minimum occurs for $\phi = \phi_0$ 
and the potential is flat at the origin if $\phi_0 = \sqrt{2}$. Here we use {\it Planck-2018} data to fix the slope at the origin. 
The fact that a viable model (according to {\it Planck-2018} data) does not require $V^{\prime}=0$ at the origin makes us to abandon the idea that the potential  (\ref{sugrapot4}) originates from a phase transition from a high energy symmetric 
potential where thermal effects keep the inflaton initially at the origin. Instead we propose that it would be more natural to 
think in a potential not committed with developing inflation close to the origin. Although both potentials are equivalent 
(since they give the same observables), the potential given by Eq.~(\ref{SuPo1}) and Fig.~\ref{SugraPot2} does not suggest 
that the origin is privileged in any way being inflation a transient phenomenon where both the beginning and end to inflation 
can be found and a {\it total} number of $e$-folds can be quantified. For calculations, however, we find more convenient to 
work with Eq.~(\ref{sugrapot4}) and we stick to this presentation of the potential in what follows.}
\label{SugraPot3}
\end{center}
\end{figure}

This potential is illustrated in Fig. \ref{SugraPot2} for some typical values of the parameters.
We see that the potential given by Eq. (\ref{SuPo1}) has a minimum at $\phi=0$ with vanishing energy. 
For $\phi=\phi_0$ the derivative of the potential is
\beq
\label{SuPoDer1}
V^{\prime}=\frac{1}{2}\Lambda^2\phi_0(8-4\phi_0^2),
\eeq
thus, the potential is also flat for $\phi_0=\sqrt{2}\,.$ To make further analytical progress we would have to find $\phi_H$ 
perhaps through an expansion of the spectral index for $\phi-\phi_0$ small. Equivalently we can can shift the origin away 
from the minimum at $\phi=0$ and make a reflection around the new origin i.e., by making the field transformation 
$\phi\rightarrow -\phi+\phi_0$. In terms of the original field $z$ we have that $z\rightarrow -z+z_0$ then 
\beq
\label{kpot}
K(z,z^*)\rightarrow z z^*+ \cdot  \cdot  \cdot \, ,
\eeq
with superpotential  \cite{Ellis:1982dg,*Nanopoulos:1982bv, *Ovrut:1983my,*Holman:1984yj,Ross:1995dq}
\beq
\label{superpot}
W(z)\rightarrow \Lambda(z-z_0)^2, 
\eeq
the resulting potential is given by
\begin{widetext}
\beq
\label{sugrapot3}
V=\Lambda^2 e^{z z^*}(z-z_0)(z^*-z^*_0) \left[ 4+z^*(z+z_0+z(z-z_0)z^*)+(z-3z_0-z^2z^*+zz^*z_0)z^*_0\right].
\eeq
\end{widetext}
Parameterising as in Eq.~(\ref{complexfield}) we set the stable direction $\chi=\chi_0=0$ and study the potential along the 
inflationary $\phi$-direction only \cite{Ellis:1982dg,*Nanopoulos:1982bv, *Ovrut:1983my,*Holman:1984yj,Ross:1995dq, German:2004vf}
{\small
\beq
\label{sugrapot4}
V=\Lambda^2 e^{\frac{1}{2}\phi^2}(\phi-\phi_0)^2 \left[ 2+\frac{1}{8}(\phi-\phi_0)\left(6\phi_0+\phi(2+\phi^2-\phi\phi_0)\right)\right].
\eeq}
This potential is illustrated in Fig. \ref{SugraPot3} for some typical values of the parameters. This form of the potential can be viewed as an example of \textit{inflection point inflation}. Such type of models have been studied in e.g.,  \cite{BuenoSanchez:2006rze,Allahverdi:2006we}, in the context of the Minimal Supersymetric Standard Model (MSSN). There, the particular $A$-term in the inflationary potential (also known as $A-$term inflation \cite{Lyth:2006ec}) can induce a saddle point at which $V^{\prime}(\phi=0)=V^{\prime\prime}(\phi=0)=0$. While this fails to reproduce an acceptable $n_s$ \cite{Allahverdi:2006iq,Gomes:2018uhv}, a point of inflection with non-zero $V^{\prime}(\phi=0)$ can successfully do so, as we show explicitly for our case in Sec.~\ref{version}. 

In our case, we see that the potential of Fig.~\ref{SugraPot3} is exactly of the same shape as the one shown by Fig.~\ref{SugraPot2} but looked at from a different frame. The potential has not been deformed, i.e., it has only been displaced and reflected. 
Thus, all observables calculated from the potential Eq.~(\ref{sugrapot4}) should take exactly the same values as the ones 
calculated from Eq.~(\ref{SuPo1}).
The new SR parameters have the same values when evaluated at the new $\phi_H$  as the old SR parameters 
had when evaluated at the old $\phi_H$.
This shows explicitly how physically relevant quantities like the spectral index, tensor to scalar ratio and all other observables 
do not depend on the frame in which the potential is evaluated. These features are discussed in detail in the next section.

\section{Slow-roll parameters and observables} \label{Slow Roll} 

The potential an all its even-number of derivatives are unchanged by the operations of translation and reflection but odd-number 
of derivatives of the potential change sign (because $\phi \rightarrow -\phi$ was used). However the SR parameters remain 
unchanged since they involve an even number of factors with odd-number of derivatives of the potential.
In slow-roll inflation, the spectral indices are given in terms of the slow-roll parameters of the model, these are given by
 (see e.g. \cite{Liddle:1994dx, Lyth:1998xn,*Liddle:2000cg})
\begin{multline}
\label{Slowparameters}
\epsilon \equiv \frac{M^{2}}{2}\left( \frac{V^{\prime }}{V }\right) ^{2},\quad
\eta \equiv M^{2}\frac{V^{\prime \prime }}{V},  \\
  \xi_2 \equiv M^{4}\frac{V^{\prime }V^{\prime \prime \prime }}{V^{2}},\quad
\xi_3 \equiv M^{6}\frac{V^{\prime 2 }V^{\prime \prime \prime \prime }}{V^{3}}.
\end{multline}

Here primes denote derivatives with respect to the inflaton $\phi$ and $M$ is the
reduced Planck mass $M=2.44\times 10^{18} \,\mathrm{GeV}$ which we set
$M=1$ in most of what follows. For the case at hand all the SR parameters are given by complicated expressions which are 
functions of the potential $V(\phi)$ and its derivatives an are not reproduced here. Thus, the SR parameters are given in terms 
of the potential {$V(\phi)$} and its derivatives. Because when going from the old potential Eq.~(\ref{SuPo1}) to the new one  
Eq.~(\ref{sugrapot4}) the potential  has not been locally changed and because odd number of derivative terms appear always 
in even numbers in the SR parameters then it follows that the observables remain unchanged.
The potential as given by Eq.~(\ref{SuPo1}) and Fig.~\ref{SugraPot2} seems more natural than that of Eq.~(\ref{sugrapot4}) 
and Fig.~\ref{SugraPot3} since it does not suggest that inflation is somehow connected with a previous phase transition from 
a symmetric phase at high energy where the inflaton is originally kept by thermal effects at the origin. 

The primordial power spectra {$\mathcal{P}_i$,} a power-law parameterized spectra of scalar and tensor perturbations, 
is usually given in terms of the spectral amplitude $A_i$ and the spectral {indices} $n_i$, where the subindex ${}_i$ 
refers to either scalar $(s)$ or tensor $(t)$ components
\begin{eqnarray}
\mathcal{P}_s(k)&=&A_s \left( \frac{k}{k_H}\right)^{(n_s-1) },\label{ss1}\\
\mathcal{P}_t(k)&=&A_t \left( \frac{k}{k_H}\right)^{n_t}=rA_s\left( \frac{k}{k_H}\right)^{n_t},
\label{ts1}
\end{eqnarray}
where $k$ is the wave number and $r\equiv \mathcal{P}_t(k)/\mathcal{P}_s(k)$ the ratio of tensor to scalar perturbations at
pivot scale $k=k_H$, here set to   $k_H = 0.002 \mathrm{Mpc}^{-1}$  \cite{Vazquez18, Planck:2013jfk,Ade:2015lrj}.
Slow-roll inflation predicts the spectrum of curvature perturbations to be close to scale-invariant 
(for a review of cosmic inflation and its relationship with cosmological observations see \cite{Vazquez18}). 
This allows a simpler parameterization of the spectra in terms of quantities evaluated at $k_H$ such as the spectral index, 
the running, and even the 
running of the running of scalar perturbations \cite{Powell:2012xz}
\begin{eqnarray}
\mathcal{P}_s(k)&=&A_s \left( \frac{k}{k_H}\right)^{(n_s-1) + \frac{1}{2}n_{sk}  \ln\left(\frac{k}{k_H}\right) + 
	       \frac{1}{6} n_{skk} \left(\ln\left(\frac{k}{k_H}\right)\right)^2  },\label{ss2}\\
\mathcal{P}_t(k)&=&A_t \left( \frac{k}{k_H}\right)^{n_t+ \frac{1}{2} n_{tk} \ln\left(\frac{k}{k_H}\right) }\label{ts2},
\end{eqnarray}
where $n_{tk}$ is the running of the tensor index $n_{t}$, $n_{sk}$ the running of the scalar index $n_{s}$  
in a self-explanatory notation. 
In the slow-roll approximation these observables are given by (see e.g. \cite{Liddle:1994dx, Lyth:1998xn, *Liddle:2000cg, Planck:2013jfk, Ade:2015lrj})
\begin{eqnarray}
n_{t} &=&-2\epsilon =-\frac{r}{8} , \label{Int} \\
n_{s} &=&1+2\eta -6\epsilon ,  \label{Ins} \\
n_{tk} &=&\frac{d n_{t}}{d \ln k}=4\epsilon\left( \eta -2\epsilon\right) = \frac{r}{64}\left(r -8(1-n_s)\right), \label{Intk} \\
n_{sk} &=&\frac{d n_{s}}{d \ln k}=16\epsilon \eta -24\epsilon ^{2}-2\xi_2, \label{Insk} \\
A_s(k) &=&\frac{1}{24\pi ^{2}} \frac{\Lambda^4}{%
\epsilon _H}, 
\label{IA} 
\end{eqnarray}
where the amplitude of density perturbations at wave number $k$ is $A_s(k)$ and the scale of 
inflation is $\Lambda$, with $\Lambda \equiv V_{H}^{1/4}$. 

\section{Choosing a convenient version of the model} \label{version} 

For calculations we find more convenient to work with Eq.~(\ref{sugrapot4}) and we stick to this presentation of the potential in 
what follows. We see that the potential given by Eq.~(\ref{sugrapot4}) has now a minimum at $\phi_0$ with vanishing energy. 
We redefine $\phi_0$ in terms of a new parameter $s$ as follows
\beq
\label{slope}
\phi_0=s/8+\sqrt {2},
\eeq
and calculate the derivative of $V$ at the origin
\beq
\label{firstderivativeV}
V^{\prime}(\phi=0)/\Lambda^2= s+\frac{3s^2}{16\sqrt{2}}+\frac{s^3}{256}.
\eeq
Thus $s$ measures the slope of the potential at the origin. Previous works \cite{Ellis:1982dg,*Nanopoulos:1982bv, *Ovrut:1983my,*Holman:1984yj,Ross:1995dq} take $s=0$  giving a 
model presently ruled out by the data \cite{Gomes:2018uhv}. Given that there is no special reason (apart from simplicity) to 
fix $V^{\prime}(\phi) = 0$ 
at the origin. We perform a Bayesian parameter fitting to determine the value of $s$ (and consequently $\phi_0$), 
consistent with {\it Planck 2018} data set. In this way the value of $\phi_0$ turns out to be very close to $\sqrt{2}$  resulting 
in a viable model.
The simplifying  assumption $V^{\prime}(\phi=0) = 0$ is typically motivated by the suggestion that thermal effects at higher energy 
put the inflaton at the origin and then a phase transition makes the inflaton rolls to its global minimum. This view is partially 
motivated by the privileged position given to the origin. In the equivalent potential of Eq.~(\ref{SuPo1}) and Fig. \ref{SugraPot2} 
no such interpretation seems to arise because the origin does not play any particular rol during inflation. We take the view, 
in any case, that the scalar field starts rolling from high energies to the global minimum of the potential and that an epoch of 
inflation occurs without any need for the inflaton to start its rolling from a privileged point. Thus, in this scheme, inflation is a 
transient phenomenon which consequently lasts
 a measurable finite ${\it total}$ number of $e$-folds. In this case both the start and the end of inflation localized in field space 
 by the conditions $| \eta | = 1$.
Having said so, the potential and  its observables can be studied using the frame which results more convenient for the 
calculation at hand keeping in mind that the displacement away from the origin is irrelevant. Thus, in what follows we study 
the potential as given by Eq.~(\ref{sugrapot4}) and we extract an approximate formula for $\phi_{H}$.
At $\phi=0$ the derivative of the potential Ec.~(\ref{sugrapot4}) is given by $V^{\prime}\left (\phi=0\right)=-\frac{1}{2}\Lambda^2\phi_0\left (8-4 \phi^2_0 \right) $ thus, $\phi_0=\sqrt{2}$ is a flat direction of the potential. As a consequence $\phi_{H}$ should be 
close to the origin and we can obtain an approximate value for it by a simple expansion of the spectral index around $\phi=0$. 
The result is
\beq
\label{fihold}
\phi_{H}^{new}\approx \frac{1}{39}\left(6\sqrt{2}-\sqrt{33+39 n_s}\right)\approx \frac{1}{12 \sqrt{2}}\left(1-n_s\right).
\eeq
Making $\phi_H \rightarrow-\phi_H+\phi_0$ we obtain an approximated expression for $\phi_{H}$ for the old potential of Eq.~(\ref{SuPo1}).
\beq
\label{fihnrw}
\phi_{H}^{old}\approx \frac{1}{39}\left(33\sqrt{2}+\sqrt{33+39 n_s}\right)\approx \frac{1}{12 \sqrt{2}}\left(23+n_s\right).
\eeq 
Using the central value reported by {\it Planck 2018} $n_s\approx 0.9649$ we find $\phi_H^{new}\approx 0.00206$ while $\phi_H^{old}\approx 1.41216$. This simple result makes explicitly clear  that the value of the field is not relevant {\it per se} since the observables 
are the same whether we use $\phi_{H}^{new}\approx 0.00206$ in the new potential or $\phi_{H}^{old}\approx 1.41216$ in the old one. 
Also, the model defined by Eq.~(\ref{sugrapot4}) and illustrated by 
Fig.~\ref{SugraPot3} has a $\Delta\phi$  of exactly the same size as in the old potential given by Eq.~(\ref{SuPo1}).
The end of inflation is here given by the condition $\eta=-1$ being $\epsilon$ much smaller than one during the whole period of inflation. 
It is convenient to use the "new" version of the potential given by Eq. (\ref{sugrapot4}) and initially assume $\phi_0=\sqrt{2}$ 
since  $\eta$ does not change appreciably for small changes in $\phi_0$. 
Also, the number of $e$-folds is not really sensitive to  small changes in $\phi$  at the end of inflation and a numerical estimate 
of $\phi_e$ yields a good approximation, however, extreme care should be taken for $\phi$ close to $\phi_H$. 
Thus, the solution to $\eta=-1$ when $\phi_0=\sqrt{2}$ is given by $\phi_e \approx 0.1694$.
For the "new" potential $\phi_{H}$ should be close to the origin and we can obtain an approximate value for it by a simple 
expansion of the spectral index around $\phi=0$. We should expect that the number of $e$-folds close to $\phi_{H}$ 
is sensitive to changes in $\phi_{H}$. Because $\phi_{H}$ depends on $\phi_0$ and $\phi_0$ weakly depends on $s$ 
we use Eq. (\ref{slope}) and also expand in $s$. The result, to first order in $s$ is
\beq
\label{fih}
\phi_{H}^{new}\approx  \frac{2-3\sqrt{2} \,s -2 n_s}{24 \sqrt{2}+9 s},
\eeq
which reduces to Eq. (\ref{fihold}) when $s=0$. We then find the parameter $s$ by requiring that the number of $e$-folds 
from $\phi_H$ to $\phi_e$  is a certain number, let us say $N=60$.
The number of $e$-folds from $\phi_H$ to the end of inflation at $\phi _e$ is given by
\begin{equation}
N\equiv -\int_{\phi _H}^{\phi_e}\frac{V({\phi })}{V^{\prime }({\phi })}{d}{\phi }.
\label{NN}
\end{equation}

\begin{table}[t]
\captionsetup{format=plain,justification=centerlast}
  \caption{Observables obtained with the value $s= -8.3\times 10^{-5}$, equivalently $\phi_0=1.414203.$ This value of $s$ 
  is first obtained from the requirement of $60$ $e$-folds of inflation using the central value for the spectral index $n_s = 0.9649$ 
  through Eqs.~(\ref{fih}) and (\ref{NN}).}
 \begin{tabular}{cccccc}
\toprule
\cline{1-6}\noalign{\smallskip}
$\phi_0$  \qquad & $N$	 \qquad	& $r$    \qquad & $n_s$   \qquad & $n_{sk}$  \qquad & $\Lambda$  (GeV) \\
\hline
\\
1.414203 \qquad & $60$ \qquad   & $8.2 \times 10^{-8}$  \qquad & $0.9649$ \qquad & $1.7 \times 10^{-3}$ \qquad & $5.5 \times 10^{14}$\\ 
\hline
\end{tabular}
\label{tab:analytical}
 \end{table}
 

We find that $s\approx -8.3\times 10^{-5}$ thus, the required slope of the potential at the origin is small indeed. The value of $s$, which controls $V^{\prime}$ at the origin (see Eq.~\eqref{firstderivativeV}), turns out to be
only a small correction to the value of $\phi_0$. As proved in the following sections, this is not unexpected as $V^{\prime}$ at the inflection point cannot be too large in order to obtain values of $n_s$ consistent with CMB observations. This fine-tuning issue of \textit{inflection point inflation} has also been discussed in the context of MSSN $A-$term inflation in \cite{BuenoSanchez:2006rze,Allahverdi:2006we}. Here we rely on the Bayesian treatment of data to determine the acceptable values of $s$. This is a relevant issue since, while the small value of $s$ does not appreciably change $\phi_H$ it is crucial in the estimation of $N$, given that $V^{\prime }({\phi })$ appears in Eq.~(\ref{NN}) is proportional to $s$.
The {\it beginning} of inflation is here given by the SR condition $\eta=1$ which occurs for $\phi_b\approx -0.09442$ giving  a {\it total} number of $e$-folds $N_T=163$. Thus, inflation is here only a transient phenomenon lasting almost thrice the required 60 $e$-folds. Observables obtained with this value for $s$ are given in Table~\ref{tab:analytical}.

\section{Analysis of the model in light of \textit{Planck 2018}  data}
\label{Planck}

Even though the primary parameters that describe the CMB spectrum
have already been  tightly constrained in several inflationary models and have little
impact on the $B$-mode spectrum, it is worthwhile to perform
a full parameter-space exploration to determine the
tensor-to-scalar ratio constraints for the model. Throughout this model, 
we assume purely
Gaussian adiabatic scalar and  tensor contributions with a flat
$\Lambda$CDM background specified by the
standard parameters (see Table~\ref{tab:posteriors}): the physical baryon ($\Omega_{\rm b} h^2$) and
cold dark matter density ($\Omega_{\rm c} h^2$) relative to the
critical density ($h$ is the  dimensionless Hubble parameter such
that $H_0=100h$ kms$^{-1}$Mpc$^{-1}$), $\theta$ being $100 \times$
the ratio of the sound horizon to angular diameter distance at
last scattering surface and $\tau$ denotes the optical depth at
reionization. We consider the tensor-to-scalar  ratio
defined previously as $r\equiv \mathcal{P}_t(k)/\mathcal{P}_s(k)$, and
hereafter we set the ratio $r=r(k_H)$ at a
scale of $k_H= 0.002 {\rm Mpc^{-1}}$. 
 The parameters describing the primordial spectrum for the model 
along with its flat priors imposed in our Bayesian analysis are shown in Table~\ref{tab:posteriors};
see \cite{Vazquez11, Vazquez12} and references therein for a Bayesian description over the primordial spectrum. 
\\

Throughout the analysis,
the  $C_\ell$'s spectra -- temperature and polarization (E \& B) -- are computed
with a modified version of the CAMB code \cite{CAMB}, and the parameter estimation is
performed using the CosmoMC  program \cite{Cosmo}.
To compute posterior probabilities for each model
we use the full-mission Planck 2015 observations of temperature and polarization anisotropies of the
CMB radiation (PLK; \cite{PLK})
and the B-mode polarization constraints from a joint analysis analysis 
of BICEP2, Keck Array, and Planck  (BKP; \cite{BKP}) data.
Furthermore, we use Baryon acoustic oscillations data  to break parameter degeneracies from CMB measurements
 (BAO; \cite{BAO} and references therein). We refer to this combined dataset as Dataset I. 
 Similarly, to incorporate the up-to-date version of the data we include  the full-mission \textit{Planck 2018 }
 (TT,TE,EE+lowE+lensing)  \cite{PLK18}, the  Keck Array and BICEP2 Collaborations 2016 \cite{BKP16} 
 and the BAO data \cite{BAO}, named as Dataset II.
 
\begin{table*}
\captionsetup{format=plain,justification=centerlast}
\small
  \caption{Parameters and prior ranges used in our analysis.
  Last four columns display mean values along with 1-$\sigma$ estimation. 
  For one-tailed distributions the upper limit 95\% CL is given. For two-tailed the 68\% is shown.
Derived parameters are labeled with~$^*$. }

\begin{tabular}{cccccc}
\toprule
\cline{1-6}\noalign{\smallskip}
 Parameter		& Prior range  & Dataset I & Dataset II & Dataset I & Dataset II\\ [0.15cm]
\hline [0.1cm]
 $\Omega_{\rm b} h^2$  	 	& [0.01, 0.03]  &  $0.02231 \pm 0.00014$ & $0.02242 \pm 0.00013$ & $0.02230 \pm0.00020$ & $0.02242 \pm 0.00018$ \\ [0.1cm]
$\Omega_{\rm DM} h^2$   	&  [0.01 , 0.3]  & $0.1185 \pm 0.0010$       &  $0.11946 \pm 0.0009$ & $0.1185 \pm 0.0015 $ & $0.1194 \pm 0.0012$\\ [0.1cm]
$\theta$               			&  [1.0 , 1.1]    & $1.04088 \pm 0.00029$   & $1.04098 \pm 0.00029$& $1.04085 \pm0.00043$ & $1.04099 \pm 0.00039$\\	[0.1cm]	    
$\tau$               				& [0.01 , 0.3]   & $0.078 \pm 0.014$           & $0.057 \pm 0.007$ & $0.078 \pm 0.020$ & $0.057 \pm0.010$  \\[0.1cm] 
\hline 
$\log[10^{10}A_{\rm s}]$  & [2.5 , 4.0]  &  $3.089 \pm 0.027$ & $3.049 \pm 0.014$ & $3.088 \pm 0.039$ & $3.050 \pm 0.020$ \\ [0.1cm]	   
$\phi_0$ [1.414 --]       & [190  , 210] &  $202 \pm 1.3 $ & $202 \pm 1.1$ & $ 202 \pm 5.6$ & $200 \pm 4.8$\\ [0.1cm]	   
$r_{02}$          			&[ 0 , 0.5]   & $< 0.078$  & $< 0.065$ & $<0.106$ & $<0.921$\\ [0.1cm]	   
$N$          			& [30 , 90]   &  $60$  &60 & $63.2 \pm 11.7$ & $58.5 \pm 8.3$\\ [0.1cm]	   
\hline
$^*n_s$          		&  -     &  $0.9670  \pm 0.0042$  & $0.9661 \pm 0.0037$ & $0.9670 \pm 0.0058$ & $0.9661 \pm 0.0052$\\ [0.1cm]	   
$^*n_{sk}$          	&  -     &$-0.0018 \pm 0.0001$ & $-0.0017\pm 0.00009$ & $-0.0017 \pm 0.0007$ & $-0.0020 \pm 0.0006$\\ [0.1cm]	   
$^*n_{tk} [10^{-10}]$ &  -     & $-3.758 \pm 0.070$ & $-3.770 \pm 0.047$ & $-4.508 \pm 3.821$ & $-5.476 \pm 3.552$\\ [0.1cm]	   
\bottomrule
\hline
\end{tabular}
\label{tab:posteriors}
 \end{table*}

\begin{figure}[htbp!]
\captionsetup{format=plain,justification=centerlast}
\par
\begin{center}
\includegraphics[trim = 0mm  0mm 1mm 1mm, clip, width=6.cm, height=5.cm]{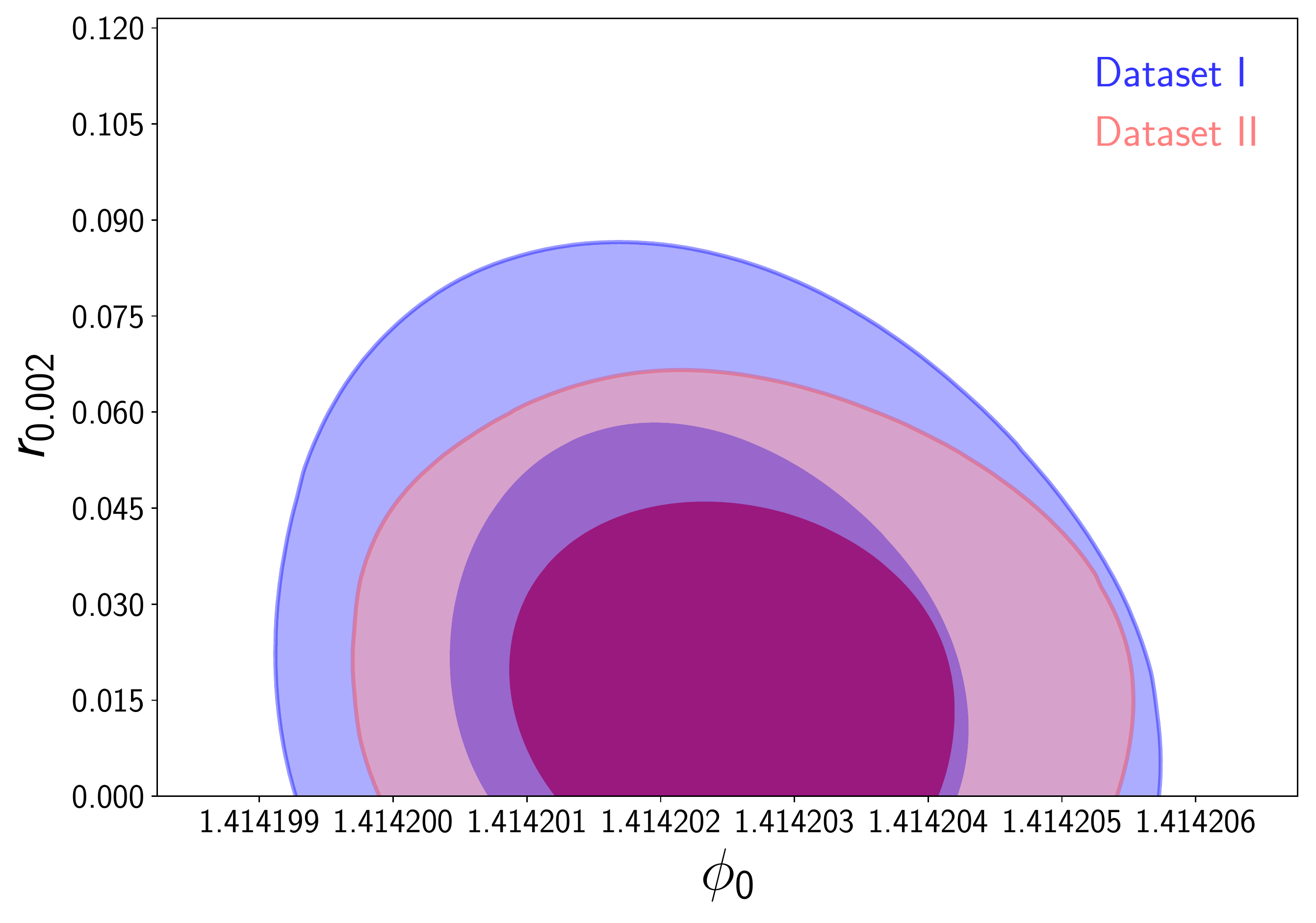}
\includegraphics[trim = 0mm  0mm 1mm 1mm, clip, width=6.cm, height=5.cm]{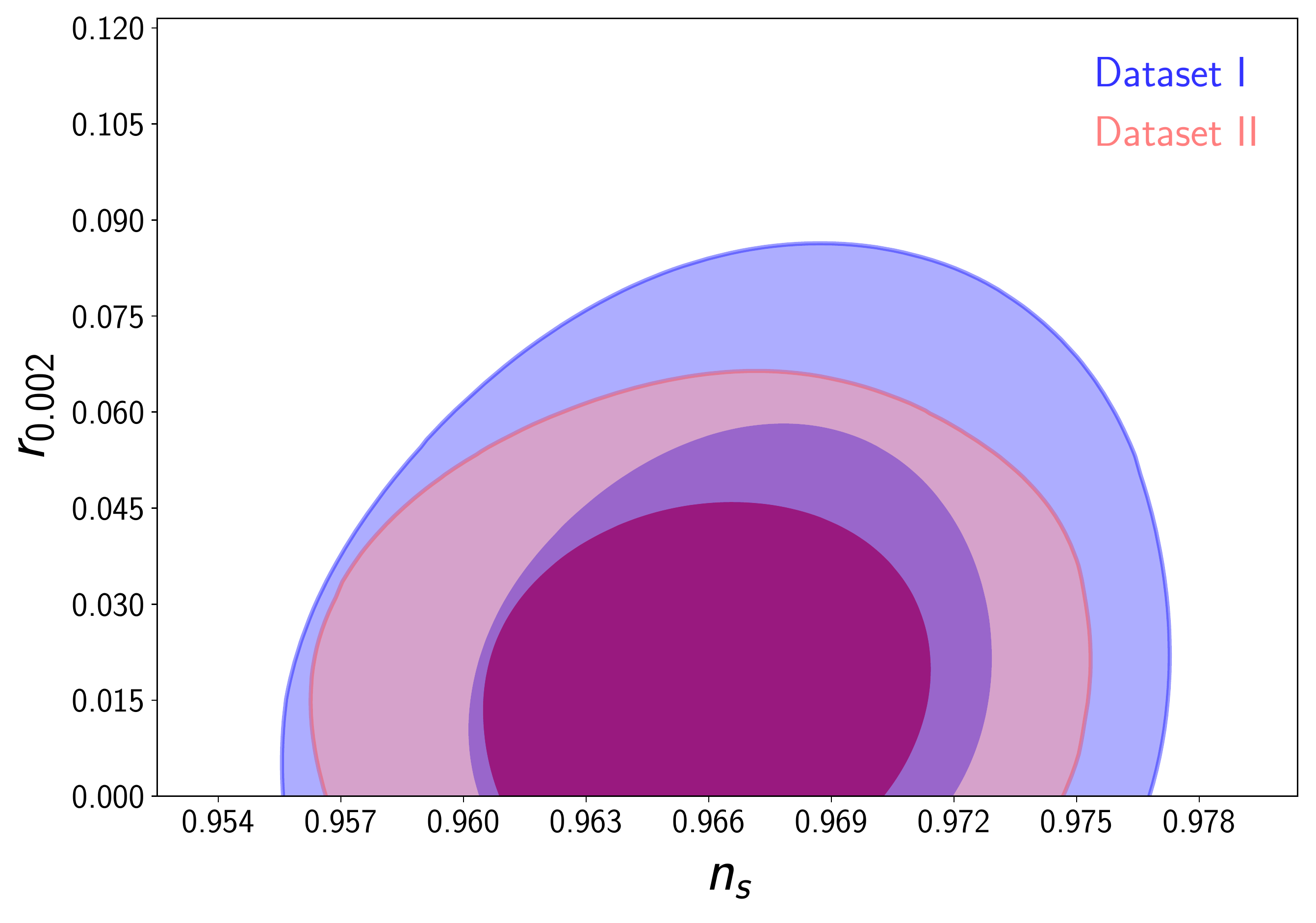}
\end{center}
\caption{2D probability posterior distributions for the power spectrum Sugra  model parameters ($N=60$); 
2D constraints are plotted with $1\sigma$ and $2\sigma$ confidence contours. 
}
\label{fig:SG}
\end{figure}

\vspace{0.75cm}

\begin{figure}[t]
\captionsetup{format=plain,justification=centerlast}
\par
\begin{center}
\includegraphics[trim = 0mm  0mm 1mm 1mm, clip, width=6.cm, height=5.cm]{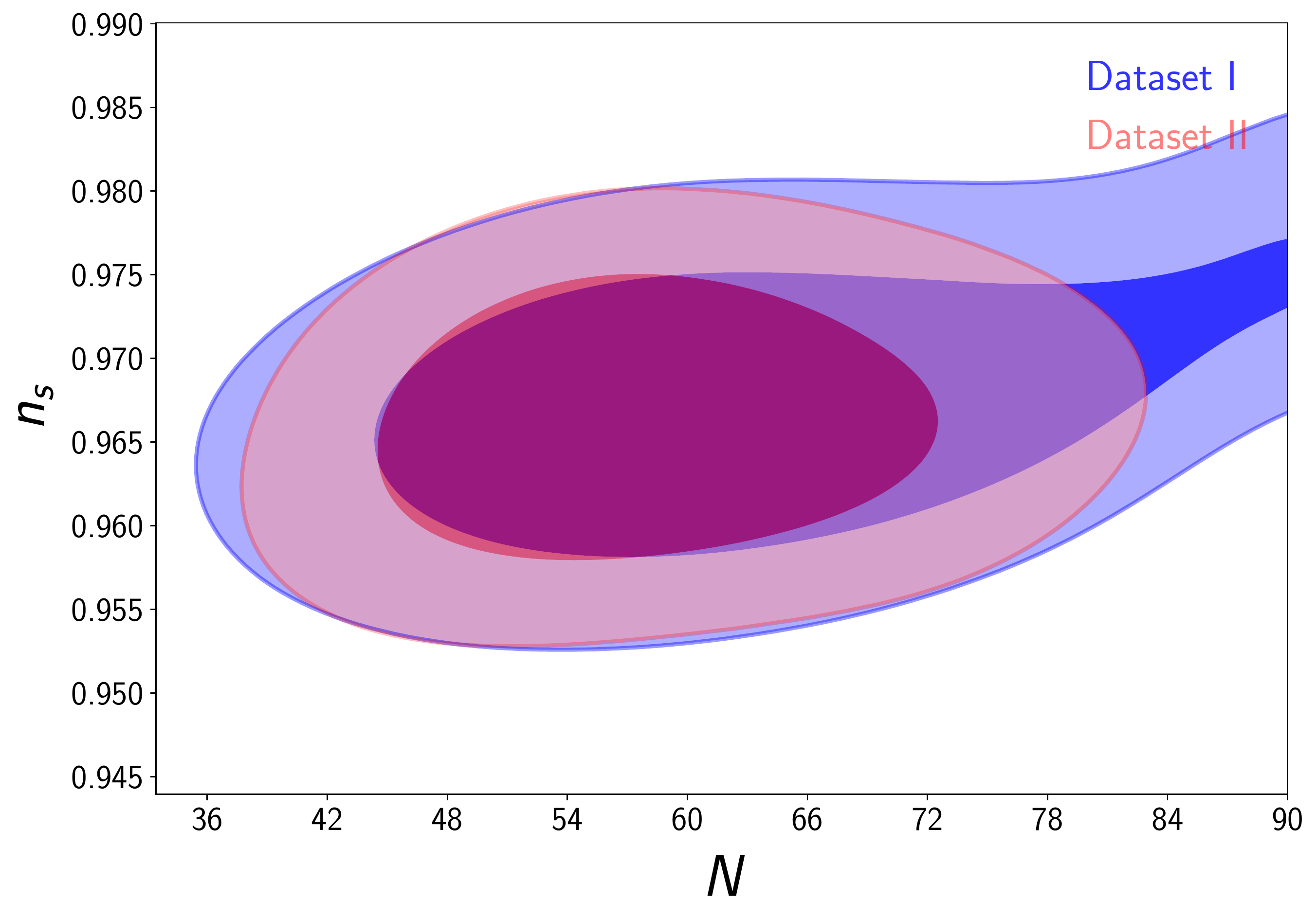}
\includegraphics[trim = 0mm  0mm 1mm 1mm, clip, width=6.cm, height=5.cm]{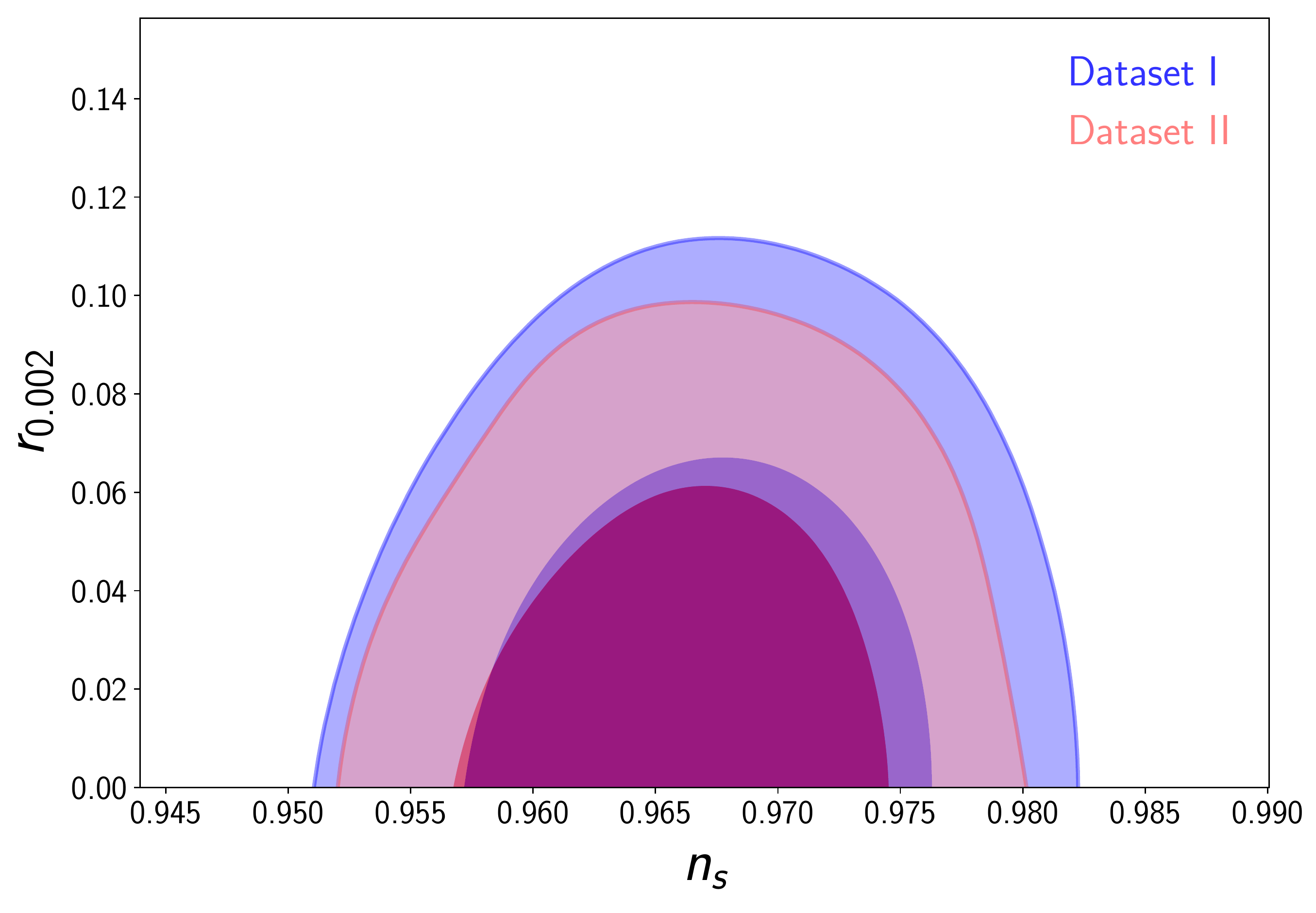}
\end{center}
\caption{2D probability posterior distributions for the power spectrum Sugra  model parameters 
letting the $e$-fold number be a free parameter. 
2D constraints are plotted with $1\sigma$ and $2\sigma$ confidence contours.
}
\label{fig:SGN}
\end{figure}

 Table \ref{tab:posteriors} summarizes the observational constraints of the parameters
that describe the model (as well as the derived parameters, labelled by $^*$).
Figure \ref{fig:SG} displays 2D marginalized posterior distributions 
of the parameters used to describe the  model along to Dataset I (blue) and Dataset II (red).
A similar analysis is performed when the $e$-fold 
number $N$ is considered as an extra free quantity (shown in Figure \ref{fig:SGN}). 
The inner ellipses show the 68\% confidence region, and the outer edges the 95\% region. 


The Dataset II, considering $N=60$ constrains the parameter of the  model as $\phi_0 =1.414202 \pm 1.1\times 10^{-6}$ 
at 68 $\%$ C.L., and $r< 0.065$ at 95 $\%$ C.L., which can be mapped into  $n_s = 0.9661 \pm0.0037 $, in agreement with 
\textit{Planck 2018} observations, with the addition of a negative running of the spectral index $n_{sk}=-0.0017\pm 0.00009$. 
Moreover, the updated information allows to tighten the constraints on the $e$-fold
parameter such as $N= 58.5 \pm 8.3$ in agreement with the initial assumptions made  in Table~\ref{tab:analytical}.
\\

\section{Discussion of Results and Conclusions} \label{Conclusions}

In this paper we presented a model of inflation derived from an $\mathcal{N}=1$ supergravity realization, taking advantage 
of the stable direction and interpreting the dynamical part as the inflaton with a transit slow-roll period. We have 
exploited the mirror-shift symmetry of the slow-roll parameters (and the slow-roll inflation observables) to displace the 
vacuum away from the origin at $\phi = 0$ and represent the potential as in Eq.~(\ref{sugrapot4}), plotted in 
Fig.~\ref{SugraPot3}. Approximated, analytic expressions are easier to find here as compared with the original potential given by Eq.~(\ref{SuPo1}). Through a Bayesian parameter fitting we have determined the 
range of values for $\phi_0$ compatible with the latest {\it Planck 2018} data, as well as the predicted spectral index and tensor-to-scalar ratio. As shown in Fig.~\ref{fig:SG}, the value $\phi_0 = \sqrt{2}$ is ruled out by more than three standard deviations but we have shown that the supergravity model is flexible enough to 
accommodate values away from this reference value.

The results confirm our analytical assumption of $N=60$ consistent with \textit{Planck 2018} data; in the upper panel of Fig.~\ref{fig:SGN}, the number of $e$-folds is constrained to a finite range of values centered around $N=58.5$. This is an important feature of the model. It stems from the transient aspect of the inflationary phase. Moreover, for a given total number of $e$-folds, our inflationary model admits only specific values of the equation of state during the reheating stage, consistent with observed spectral index values \cite{German:2018wrx}.

\section*{Acknowledgements}
 
G.G. acknowledges financial support from UNAM-PAPIIT, IN104119, {\it Estudios en Gravitaci\'on y Cosmolog\'ia}. J.C.H. Acknowledges support from SEP-CONACYT grant
282569. The work of A.M. is supported by the postdoctoral grants programme of DGAPA-UNAM.
J.A.V. acknowledges the support provided by FOSEC SEP-CONACYT Investigaci\'on B\'asica  A1-S-21925, and DGAPA-PAPIIT  IA102219. FXLC thanks the Instituto de Ciencias F\'isicas at Universidad Nacional Aut\'onoma de M\'exico (ICF-UNAM) for its kind hospitality during the development of this work, and the joint support by CONACyT and DAIP-UG.

\bibliography{biblio}

\begin{thebibliography}{10}

\bibitem{PLK18}
N.~Aghanim et~al.
\newblock {Planck 2018 results. VI. Cosmological parameters}.
\newblock 2018, astro-ph/1807.06209.

\bibitem{Akrami:2018odb}
Y.~Akrami et~al.
\newblock {Planck 2018 results. X. Constraints on inflation}.
\newblock 2018, astro-ph/1807.06211.

\bibitem{Guth:1980zm}
Alan~H. Guth.
\newblock {The Inflationary Universe: A Possible Solution to the Horizon and
  Flatness Problems}.
\newblock {\em Phys. Rev.}, D23:347--356, 1981.
\newblock [Adv. Ser. Astrophys. Cosmol.3,139(1987)].

\bibitem{Martin:2013tda}
Jerome Martin, Christophe Ringeval, and Vincent Vennin.
\newblock {Encyclopædia Inflationaris}.
\newblock {\em Phys. Dark Univ.}, 5-6:75--235, 2014, astro-ph/1303.3787.

\bibitem{BKP16}
P.~A.~R. Ade et~al.
\newblock {Improved Constraints on Cosmology and Foregrounds from BICEP2 and
  Keck Array Cosmic Microwave Background Data with Inclusion of 95 GHz Band}.
\newblock {\em Phys. Rev. Lett.}, 116:031302, 2016, astro-ph/1510.09217.

\bibitem{German:2015qjq}
Gabriel Germán, Alfredo Herrera-Aguilar, Juan~Carlos Hidalgo, and Roberto~A.
  Sussman.
\newblock {Canonical single field slow-roll inflation with a non-monotonic
  tensor-to-scalar ratio}.
\newblock {\em JCAP}, 1605(05):025, 2016, astro-ph/1512.03105.

\bibitem{Ellis:1982dg}
John~R. Ellis, Dimitri~V. Nanopoulos, Keith~A. Olive, and K.~Tamvakis.
\newblock {Fluctuations in a Supersymmetric Inflationary Universe}.
\newblock {\em Phys. Lett.}, 120B:331--334, 1983.

\bibitem{Nanopoulos:1982bv}
Dimitri~V. Nanopoulos, Keith~A. Olive, M.~Srednicki, and K.~Tamvakis.
\newblock {Primordial Inflation in Simple Supergravity}.
\newblock {\em Phys. Lett.}, 123B:41--44, 1983.

\bibitem{Ovrut:1983my}
Burt~A. Ovrut and Paul~J. Steinhardt.
\newblock {Supersymmetry and Inflation: A New Approach}.
\newblock {\em Phys. Lett.}, 133B:161--168, 1983.

\bibitem{Holman:1984yj}
R.~Holman, Pierre Ramond, and Graham~G. Ross.
\newblock {Supersymmetric Inflationary Cosmology}.
\newblock {\em Phys. Lett.}, 137B:343--347, 1984.

\bibitem{Ross:1995dq}
Graham~G. Ross and Subir Sarkar.
\newblock {Successful supersymmetric inflation}.
\newblock {\em Nucl. Phys.}, B461:597--624, 1996, hep-ph/9506283.

\bibitem{Cremmer:1982en}
E.~Cremmer, S.~Ferrara, L.~Girardello, and Antoine Van~Proeyen.
\newblock {Yang-Mills Theories with Local Supersymmetry: Lagrangian,
  Transformation Laws and SuperHiggs Effect}.
\newblock {\em Nucl. Phys.}, B212:413, 1983.
\newblock [413(1982)].

\bibitem{German:2004vf}
G.~German and Axel de~la Macorra.
\newblock {A Model of inflation independent of the initial conditions, with
  bounded number of e-folds and n(s) larger or smaller than one}.
\newblock {\em Phys. Rev.}, D70:103521, 2004, astro-ph/0410133.

\bibitem{BuenoSanchez:2006rze}
J.~C. Bueno~Sanchez, Konstantinos Dimopoulos, and David~H. Lyth.
\newblock {A-term inflation and the MSSM}.
\newblock {\em JCAP}, 0701:015, 2007, hep-ph/0608299.

\bibitem{Allahverdi:2006we}
Rouzbeh Allahverdi, Kari Enqvist, Juan Garcia-Bellido, Asko Jokinen, and Anupam
  Mazumdar.
\newblock {MSSM flat direction inflation: Slow roll, stability, fine tunning
  and reheating}.
\newblock {\em JCAP}, 0706:019, 2007, hep-ph/0610134.

\bibitem{Lyth:2006ec}
David~H. Lyth.
\newblock {MSSM inflation}.
\newblock {\em JCAP}, 0704:006, 2007, hep-ph/0605283.

\bibitem{Allahverdi:2006iq}
Rouzbeh Allahverdi, Kari Enqvist, Juan Garcia-Bellido, and Anupam Mazumdar.
\newblock {Gauge invariant MSSM inflaton}.
\newblock {\em Phys. Rev. Lett.}, 97:191304, 2006, hep-ph/0605035.

\bibitem{Gomes:2018uhv}
Cláudio Gomes, Orfeu Bertolami, and João~G. Rosa.
\newblock {Inflation with $Planck$ data: A survey of some exotic inflationary
  models}.
\newblock {\em Phys. Rev.}, D97(10):104061, 2018, hep-th/1803.08084.

\bibitem{Liddle:1994dx}
Andrew~R. Liddle, Paul Parsons, and John~D. Barrow.
\newblock {Formalizing the slow roll approximation in inflation}.
\newblock {\em Phys. Rev.}, D50:7222--7232, 1994, astro-ph/9408015.

\bibitem{Lyth:1998xn}
David~H. Lyth and Antonio Riotto.
\newblock {Particle physics models of inflation and the cosmological density
  perturbation}.
\newblock {\em Phys. Rept.}, 314:1--146, 1999, hep-ph/9807278.

\bibitem{Liddle:2000cg}
Andrew~R. Liddle and D.~H. Lyth.
\newblock {\em {Cosmological inflation and large scale structure}}.
\newblock 2000.

\bibitem{Vazquez18}
J.~Alberto Vázquez, Luis~E. Padilla, and Tonatiuh Matos.
\newblock {Inflationary Cosmology: From Theory to Observations}.
\newblock 2018, astro-ph/1810.09934.

\bibitem{Planck:2013jfk}
P.~A.~R. Ade et~al.
\newblock {Planck 2013 results. XXII. Constraints on inflation}.
\newblock {\em Astron. Astrophys.}, 571:A22, 2014, astro-ph/1303.5082.

\bibitem{Ade:2015lrj}
P.~A.~R. Ade et~al.
\newblock {Planck 2015 results. XX. Constraints on inflation}.
\newblock {\em Astron. Astrophys.}, 594:A20, 2016, astro-ph/1502.02114.

\bibitem{Powell:2012xz}
Brian~A. Powell.
\newblock {Scalar runnings and a test of slow roll from CMB distortions}.
\newblock 2012, astro-ph/1209.2024.

\bibitem{Vazquez11}
J.~Alberto Vazquez, A.~N. Lasenby, M.~Bridges, and M.~P. Hobson.
\newblock {A Bayesian study of the primordial power spectrum from a novel
  closed universe model}.
\newblock {\em Mon. Not. Roy. Astron. Soc.}, 422:1948--1956, 2012,
  astro-ph/1103.4619.

\bibitem{Vazquez12}
J.~Alberto Vazquez, M.~Bridges, M.~P. Hobson, and A.~N. Lasenby.
\newblock {Model selection applied to reconstruction of the Primordial Power
  Spectrum}.
\newblock {\em JCAP}, 1206:006, 2012, astro-ph/1203.1252.

\bibitem{CAMB}
Antony Lewis, Anthony Challinor, and Anthony Lasenby.
\newblock {Efficient computation of CMB anisotropies in closed FRW models}.
\newblock {\em Astrophys. J.}, 538:473--476, 2000, astro-ph/9911177.

\bibitem{Cosmo}
Antony Lewis and Sarah Bridle.
\newblock {Cosmological parameters from CMB and other data: A Monte Carlo
  approach}.
\newblock {\em Phys. Rev.}, D66:103511, 2002, astro-ph/0205436.

\bibitem{PLK}
P.~A.~R. Ade et~al.
\newblock {Planck 2015 results. XIII. Cosmological parameters}.
\newblock {\em Astron. Astrophys.}, 594:A13, 2016, astro-ph/1502.01589.

\bibitem{BKP}
P.~A.~R. Ade et~al.
\newblock {Joint Analysis of BICEP2/$Keck Array$ and $Planck$ Data}.
\newblock {\em Phys. Rev. Lett.}, 114:101301, 2015, astro-ph/1502.00612.

\bibitem{BAO}
Lauren Anderson et~al.
\newblock {The clustering of galaxies in the SDSS-III Baryon Oscillation
  Spectroscopic Survey: baryon acoustic oscillations in the Data Releases 10
  and 11 Galaxy samples}.
\newblock {\em Mon. Not. Roy. Astron. Soc.}, 441(1):24--62, 2014,
  astro-ph/1312.4877.

\bibitem{German:2018wrx}
Gabriel German, Juan~Carlos Hidalgo, and Ariadna Montiel.
\newblock Work in progress, 2019.

\end{thebibliography}
\bibliographystyle{hunsrt}

\end{document}